\documentclass[twocolumn,prb,aps]{revtex4-2}
\usepackage{dcolumn}
\usepackage{amsmath}
\usepackage{amsfonts}
\usepackage{amssymb}
\usepackage{calligra}
\usepackage{bm}
\usepackage{graphicx}
\usepackage{float}
\usepackage{subfigure}
\usepackage{natbib}
\usepackage{physics}
\usepackage[]{hyperref}
\usepackage{wasysym}
\begin{document}

\title{Directional Localization in Disordered 2D Tight-Binding Systems: Insights from Single Particle Entanglement Measures}

\author{Mohammad Pouranvari}
\email{m.pouranvari@umz.ac.ir}
\affiliation{Department of Solid-State Physics, Faculty of Science,
University of Mazandaran, Babolsar, Iran.}

\date{\today}

\begin{abstract}
We investigate the directional localization properties of wave-functions in a two-dimensional tight-binding model with uniform hopping and correlated random on-site energies. By controlling the disorder correlation strength with a parameter $\alpha$, we explore the effects of disorder on wave-function localization using Single Particle Entanglement Entropy (SPEE) and Single Particle R\'enyi Entropy (SPRE) at different values of $q$. Our analysis includes two distinct randomness structures: row-wise and fully correlated disorder. We find that row-wise disorder maintains maximal entanglement for horizontal cuts while enhancing horizontal spread for vertical cuts as $\alpha$ increases. In contrast, fully correlated disorder leads to reduced vertical entanglement for horizontal cuts and increased horizontal entanglement for vertical cuts with rising $\alpha$. Additionally, our results show that the difference between SPEE and SPRE provides valuable insights into localization behavior. These findings highlight the significance of directional properties in understanding localization transitions in disordered systems.
\end{abstract}

\maketitle

\section{Introduction} \label{sec:introduction}

Disorder and localization in condensed matter systems are pivotal topics in understanding various physical phenomena. Understanding the localization properties and spatial distribution of wave-functions in disordered systems is a fundamental problem in condensed matter physics\cite{billy2008direct, vojta2019disorder, scalettar1991localization, sanchez2010disordered}. Anderson localization, first introduced by P.W. Anderson in 1958, describes the absence of diffusion of electronic wave-functions in disordered media\cite{PhysRev.109.1492}. This phenomenon occurs when disorder in a system is sufficiently strong to cause the exponential localization of wave-functions, preventing electronic transport\cite{lagendijk2009fifty}. Anderson localization is a cornerstone in the study of disordered systems, providing a fundamental understanding of how disorder impacts the behavior of wave-functions. It has profound implications for various physical systems, including semiconductors, cold atoms, and photonic materials, and serves as a crucial theoretical framework for investigating localization transitions\cite{markos2006numerical, gade1993anderson, giamarchi1988anderson, aspect2009anderson, storzer2006observation, bulaevskii1985anderson, casacuberta1995anderson, weaver1990anderson, schwartz2007transport}.

Several measures are commonly used to investigate the localization and delocalization properties of systems. Conductivity is a fundamental measure, providing direct information about the transport properties of electrons and indicating whether a system is in a metallic (delocalized) or insulating (localized) phase\cite{abrahams1980non, czycholl1981conductivity, kolovsky2024effects, yutushui2024localization}. The localization length, which quantifies the spatial extent of localized states, is another key measure, particularly in disordered systems where it characterizes how wave-functions decay\cite{mackinnon1981one, furlan1998electronic, gasparian2009localization, del2012localization, mondal2020relation}. Transmission, often studied in mesoscopic systems, provides insights into how efficiently waves or particles traverse a system, with lower transmission indicating stronger localization\cite{shi2012transmission, asatryan2012transmission, monthus2009statistics, gredeskul2012anderson, ossipov2018scattering}. Entanglement measures, such as entanglement entropy, have gained prominence for their ability to capture the spread and complexity of quantum states in a way that complements traditional measures\cite{ippoliti2021entanglement, osterloh2002scaling, osborne2002entanglement, somma2004nature, vidal2004entanglement, chakravarty2010scaling}. The participation ratio (and its related quantity the multi-fractal properties), which quantifies the number of sites significantly occupied by a wave-function, is also widely used to distinguish between localized and extended states\cite{rodriguez2011multifractal, brandes1996critical, PhysRevB.99.155121, lindinger2017multifractal, ujfalusi2015finite, burmistrov2013multifractality}. Each of these measures contributes to a multifaceted understanding of localization phenomena in both classical and quantum systems.

Various measures have been developed to quantify entanglement in quantum systems, each offering unique insights into the nature of quantum correlations. Entanglement entropy, including von Neumann entropy and R\'enyi entropy, is a widely used measure that captures the degree of entanglement between subsystems by analyzing the entropy of reduced density matrices\cite{eisert2010colloquium, calabrese2005evolution,https://doi.org/10.1155/2015/397630, PhysRevB.100.195109,  islam2015measuring, yao2010entanglement, swingle2010entanglement, PhysRevB.96.045123, zhao2013finite, osborne2002entanglement, jia2008entanglement, doi:10.1142/S0217732324500032}. Negativity is another important measure, particularly useful in mixed-state entanglement, which quantifies the extent to which a state's partial transpose has negative eigenvalues\cite{turkeshi2020negativity, merritt2023entanglement}. Concurrence (C) is commonly used to measure entanglement in bipartite quantum systems, providing a straightforward method to determine the entanglement of two qubits\cite{qian2005characterizing, najarbashi2017quantum, koh2014entanglement, dutta2010quantum}. Other measures include the entanglement of formation\cite{wootters2001entanglement, wolf2004gaussian, chen2005entanglement}, which quantifies the resources needed to create a given entangled state, and the mutual information\cite{kraskov2004estimating, batina2011mutual, hutter2001distribution, steuer2002mutual}, which measures the total correlations between subsystems. Each of these measures provides a different perspective on entanglement, enabling a comprehensive understanding of quantum correlations in various physical systems.

The spatial-localization properties of quantum states in disordered systems have been a subject of extensive research, leading to significant insights into the nature of wave functions and their corresponding entropies. Pipek and Varga (1992) introduced a classification scheme for one-particle states based on a localization quantity termed structural entropy, establishing a universal relation between this entropy and the delocalization index, or participation ratio, independent of the lattice geometry and size \cite{pipek1992universal}. This foundational work paved the way for further investigations into the mathematical characterization of localized and fractal distributions in extended systems, as explored by Pipek and Varga (1994), where they developed a shape-analysis method for wave functions that can be applied even in complex systems lacking regular geometrical structures \cite{https://doi.org/10.1002/qua.560510619}.

Varga and Pipek (2003) advanced this framework by utilizing R\'enyi entropies to characterize the shape and extension of phase space representations of quantum wave functions in disordered systems, demonstrating the applicability of their formalism through numerical simulations \cite{PhysRevE.68.026202}. Additionally, their exploration of power-law localization at the metal-insulator transition highlighted the complexities of wave functions in quasi-periodic potentials, which contradicted previous assumptions regarding exponential localization \cite{PhysRevB.46.4978}. Further research by Varga and Pipek (1998) delved into generalized localization lengths in one-dimensional systems with correlated disorder, indicating that power-law localization may not be dismissed even in cases previously thought to exhibit exponential decay \cite{ImreVarga_1998}. More recently, Varga and M\'ndez-Berm\'udez (2008) examined entanglement in disordered systems at criticality, revealing scaling properties in various entanglement measures in the presence of disorder, further emphasizing the intricate relationships between localization and entanglement in quantum systems \cite{https://doi.org/10.1002/pssc.200777589}.

In addition, previous studies have extensively used entanglement measures, such as SPEE and SPRE (see below for explanation), to characterize wave-functions. These measures provide insights into the complexity and spread of wave-functions beyond what traditional measures like the participation ratio can offer. Specifically, SPEE and SPRE capture the degree of localization and delocalization in quantum states, offering a more nuanced understanding of wave-function behavior.

Despite these advancements, there is a research gap in understanding the directional preferences of wave-function spread in two-dimensional (2D) systems with disorder. Traditional measures like the participation ratio do not adequately distinguish between wave-functions that are localized or extended in specific directions. This limitation underscores the need for more sensitive and directional-specific measures. Specifically, these entanglement measures allow us to better understand the following: First, identify if the wave-function is more extended along the horizontal or vertical direction by comparing the entropies from the respective cuts. Second, investigate anisotropic localization properties in disordered systems, where the wave-function might localize more in one direction than the other. Finally, they provide a more nuanced characterization of the wave-function's spatial distribution, complementing the information obtained from the participation ratio.

In this work, we investigate a two-dimensional tight-binding Hamiltonian on a square lattice with uniform hopping and correlated random on-site energies. The correlation strength of the randomness is controlled by a parameter $\alpha$, which allows for a tunable degree of disorder. By calculating SPEE and SPRE for both horizontal and vertical cuts, we can detect directional preferences in the wave-function's spread.

We use two different randomness structures: row-wise disorder and fully correlated disorder. For each type, we calculate SPEE and SPRE and analyze their behavior as the disorder parameter $\alpha$ varies. Our results show distinct patterns in wave-function localization and extension, emphasizing the importance of cutting directions in these analyses. The contrasting behaviors observed in horizontal and vertical cuts underscore the importance of considering directional properties in the study of localization and entanglement in disordered systems.

Our main findings reveal significant directional localization of wave-functions in disordered systems. The behavior of SPEE and SPRE under different disorder types demonstrates that wave-functions exhibit distinct directional preferences in their spread. This research underscores the necessity of directional analysis to fully comprehend the impact of disorder on wave-function localization, providing deeper insights into anisotropic localization properties.

The remainder of this paper is structured as follows:  Section  \ref{sec:model} describes the model and methodology, including the different randomness structures investigated. Section \ref{sec:entropies} explains the Single Particle Entanglement Entropy (SPEE) and Single Particle R\'enyi Entropy (SPRE), detailing their calculation and relevance to our study, as well as the importance of horizontal and vertical cuts. In Section \ref{sec:results}, we present our results and discussion, beginning with visualizations of the wave-function configuration at the Fermi level and followed by numerical results for SPEE and SPRE. We then conclude in Section \ref{sec:conclusion}.

\section{Model and Methodology}
\label{sec:model}

We consider a two-dimensional tight-binding Hamiltonian defined on a square lattice with constant hopping terms and random on-site energies. The Hamiltonian is given by:

\begin{equation}
H = t \sum_{\langle i,j \rangle} \left( c_i^\dagger c_j + c_j^\dagger c_i \right) + \sum_{i} \epsilon_i c_i^\dagger c_i,
\end{equation}

where $c_i^\dagger$ ($c_i$) are the creation (annihilation) operators at site $i$ and we use periodic boundary conditions. The Hamiltonian consists of two main terms. The first term in the Hamiltonian represents the hopping of particles between nearest-neighbor sites. The hopping parameter $t$ is set to $-1$, which ensures uniform hopping strength throughout the lattice. The second term accounts for the random on-site energies. The $\epsilon_i$ are random variables representing the disorder in the system.

\subsection{Randomness Structures}
We introduce disorder into the system through the random on-site energies $\epsilon_i$. The randomness is characterized by a tunable parameter $\alpha$, which controls the correlation strength between the on-site energies.

Correlated disorder has been observed experimentally in various condensed matter systems, such as in materials with impurities or structural defects that exhibit spatial correlations\cite{keen2015crystallography, kuhl2000experimental, simonov2020designing, blavats2001polymers}. This type of disorder has also been extensively studied numerically, revealing its significant impact on electronic properties and localization phenomena\cite{izrailev2012anomalous, li2011theory,  carpena2002metal, croy2011anderson, de1998delocalization, izrailev1999localization, shima2004localization, porto1999correlated, schrenk2013percolation, gastiasoro2016unconventional, dudka2016critical}. Understanding correlated disorder is crucial for developing theoretical models that accurately reflect real-world materials and for predicting new physical behaviors in disordered systems. Correlated disorder for 2D systems has been studied before, and it was found that $\alpha = 2$ is the critical value for localization. However, this work pays particular attention to the directional spread or localization of the wave-function, providing new insights into the anisotropic behavior of electronic states in such systems\cite{de2004delocalization}.

We consider two different structures for the randomness. Row-wise randomness: In the first structure, we generate $N_x$ random numbers, where $N_x$ is the number of columns in the lattice. Each row of the on-site energies is set to the same random number. This structure introduces a correlated randomness along each row. Fully correlated randomness: In the second structure, we generate $N_x \times N_y$ random numbers, where $N_y$ is the number of rows in the lattice.

For either of the cases, the random numbers are generated such that they are correlated. The on-site energies are generated to mimic a sequence following the trace of a fractional Brownian motion with a specific spectral density $S(k) \propto 1/k^\alpha$. This method utilizes the discrete Fourier transform and is given by\cite{de1998delocalization, shima2004localization}:
\begin{equation}
\epsilon_i = \sum_{k=1}^{N_s} \left[k^{-\alpha} \left(\frac{2\pi}{N_s}\right)^{(1-\alpha)} \right]^{1/2} \cos\left(\frac{2\pi i k }{N_s} + \phi_k\right),
\end{equation}
where $\phi_k$ are independent random numbers uniformly distributed between $0$ and $2\pi$. $N_s$ is the number of random on-site energies to be created. In the row-wise randomness, $N_s=N_x$ and in the fully correlated randomness $N_s=N_x\times N_y$. By adjusting the parameter $\alpha$, we can control the correlation strength between the generated on-site energies. This approach offers a more nuanced representation of disorder compared to the uniform random distribution method.

These two randomness structures allow us to investigate the effects of different types of disorder on the wave-function's spatial distribution and localization properties. By comparing the results from these two methods, we can gain insights into how correlation strength and disorder structure influence the behavior of the system.

\section{Single Particle Entanglement Entropy and R\'enyi Entropy}
\label{sec:entropies}

The study of entanglement properties in disordered systems provides crucial insights into the nature of localization transitions. In this work, we utilize two important measures of entanglement: SPEE and SPRE. These measures help quantify the entanglement of a single-particle wave-function in a bipartite system.

\subsection{Single Particle Entanglement Entropy (SPEE)}
SPEE is a specific application of the von Neumann entropy to single-particle wave-functions. It provides a measure of the entanglement between two subsystems, $A$ and $B$, of a bi-partitioned lattice system.

Given a single-particle wave-function $|\psi\rangle$ in a lattice system, the total density matrix $\rho_t$ is defined as:
\begin{equation}
    \rho_t = |\psi\rangle \langle \psi|.
\end{equation}

The wave-function $|\psi\rangle$ can be expressed as:
\begin{equation}
    |\psi\rangle = \sum_{i=1}^{N} c_i |1\rangle_i \otimes_{j \ne i} |0\rangle_j,
\end{equation}
where $|1\rangle$ and $|0\rangle$ represent the occupation number basis for the lattice system.

To obtain the reduced density matrix $\rho_A$ for subsystem $A$, we trace out the degrees of freedom of subsystem $B$:
\begin{equation}
    \rho_A = \text{tr}_B (\rho_t).
\end{equation}

The SPEE is then given by:
\begin{equation}
    \text{SPEE} = -\text{tr} (\rho_A \log \rho_A).
\end{equation}

For computational simplicity, we consider the probabilities $P_A$ and $P_B$, which represent the probabilities of finding the particle in subsystems $A$ and $B$, respectively. These probabilities are calculated as:
\begin{equation}
    P_A = \sum_{i=1}^{N_A} |c_i|^2,
\end{equation}
\begin{equation}
    P_B = \sum_{i=N_A+1}^{N} |c_i|^2 = 1 - P_A.
\end{equation}

Thus, the SPEE can be expressed as:
\begin{equation}
    \text{SPEE} = -P_A \log P_A - P_B \log P_B.
\end{equation}

\subsection{Single Particle R\'enyi Entropy (SPRE)}
The R\'enyi entropy is a generalization of the von Neumann entropy, parameterized by an order $q$, which allows for a more flexible analysis of entanglement properties. The SPRE of order $q$ is defined as:
\begin{equation}
    \text{RE}_q = \frac{1}{1-q} \log \left( P_A^q + P_B^q \right),
\end{equation}
where $P_A$ and $P_B$ are the probabilities defined in the previous section.

As $q \to 1$, the R\'enyi entropy converges to the von Neumann entropy (SPEE), making SPRE a broader measure that includes SPEE as a special case.

When the probabilities $ P_A $ and $ P_B $ are equal, i.e., $ P_A = P_B = 0.5 $, both the SPEE and the SPRE reach their maximum values. For SPEE, this corresponds to $ SPEE = -P_A \log P_A - P_B \log P_B = \ln 2 $, indicating maximal entanglement between the two subsystems. Similarly, for SPRE, when $ P_A = P_B = 0.5 $, the entropy measure depends on the chosen R\'enyi parameter $ q $, but it generally also reflects maximal entanglement. Physically, this scenario signifies that the wave-function is equally distributed across both subsystems, indicating a highly delocalized state where the particle has an equal probability of being found in either half of the system.

In a two-dimensional system, the direction in which the system is cut—horizontally or vertically—affects the interpretation of SPEE and SPRE. This is because the spatial distribution of the wave-function can vary along different directions due to anisotropies or specific disorder configurations.

\subsection{Horizontal and Vertical Cuts}
To compute SPEE and SPRE, we partition the system into two equal subsystems using horizontal and vertical cuts. Both cuts are made at the middle of the system, ensuring an equal division.

In a horizontal cut, the system is divided into top and bottom halves. The cut is made such that the upper half of the lattice forms one subsystem and the lower half forms the other. This partitioning allows us to analyze the entanglement properties across the horizontal division.

In a vertical cut, the system is divided into left and right halves. Similar to the horizontal cut, but this time the cut is made vertically. The left half of the lattice forms one subsystem and the right half forms the other. This method enables us to study the entanglement properties across the vertical division.

To gain a deeper understanding of the wave-function's spatial distribution, we analyze SPEE and SPRE by performing both horizontal and vertical cuts on the system. This approach is motivated by the need to discern the directionality of the wave-function's spread, which cannot be captured by the participation ratio alone.

While the participation ratio provides a measure of the overall extent of the wave-function, it does not differentiate between spreading along different axes. A high participation ratio could result from the wave-function spreading equally in all directions or predominantly in one direction. Thus, relying solely on the participation ratio might obscure important details about the anisotropy of the wave-function.

By calculating SPEE and SPRE for both horizontal and vertical cuts, we can detect directional preferences in the wave-function's spread. Specifically, these entanglement measures allow us to better understand the following: First, identify if the wave-function is more extended along the horizontal or vertical direction by comparing the entropies from the respective cuts. Second, investigate anisotropic localization properties in disordered systems, where the wave-function might localize more in one direction than the other. Finally, they provide a more nuanced characterization of the wave-function's spatial distribution, complementing the information obtained from the participation ratio. Thus, this dual-cut strategy enhances our ability to understand the behavior of the wave-function in different disorder regimes.

\section{Results and Discussion}
\label{sec:results}

In this section, we present the key findings from our simulations, focusing on wave-function configurations, SPEE, and SPRE. We explore how varying the correlation parameter $\alpha$ affects wave-function localization and anisotropy, providing insights into the impact of correlated disorder on the system's properties.

\subsection{Wave-function Configuration at the Fermi Level}
To gain insights into the spatial distribution of the wave-function, we first examine the absolute value of the wave-function at the Fermi level, which is set to $E_F = 0$. By plotting these configurations, we visualize how the wave-function is distributed across the lattice for different values of the correlation parameter $\alpha$. These plots reveal the localization properties and the extent of the wave-function's spread in the presence of disorder.

\begin{figure}[h]
\centering
\includegraphics[width=\linewidth]{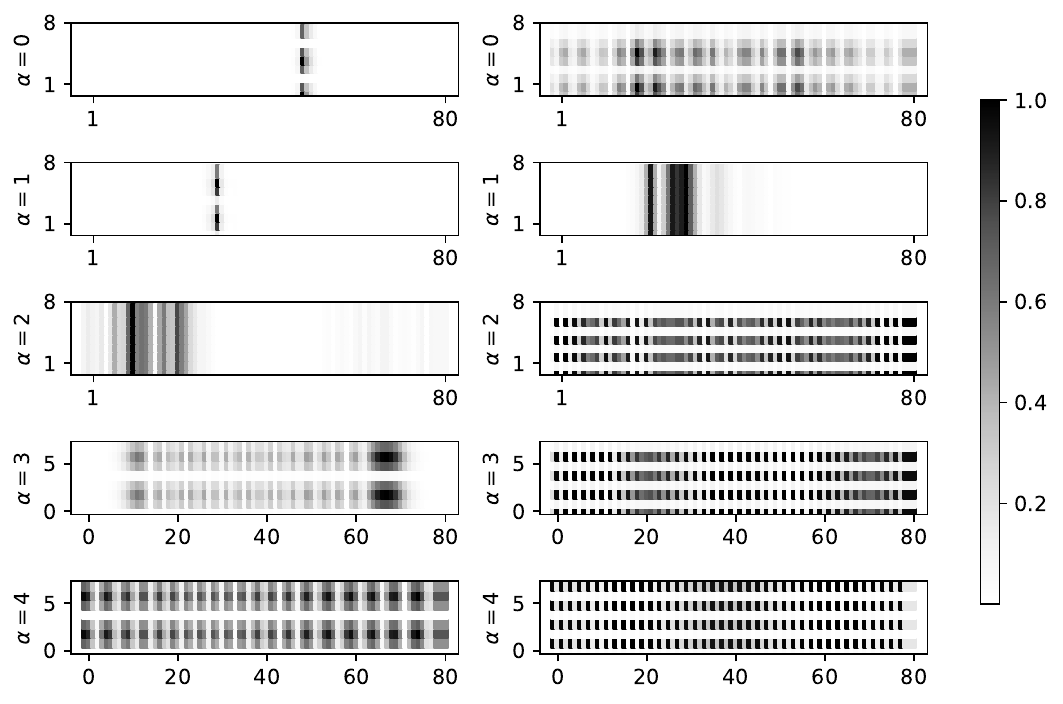}
\caption{First example of absolute value of the wave-function configurations at the Fermi level ($E_F = 0$) for five different values of $\alpha$. Left column: row-wise randomness structure. Right column: fully correlated randomness structure. The wave-function values are normalized by the maximum value of the wave-function, scaled between 0 and 1.}
\label{fig:s1}
\end{figure}

\begin{figure}[h]
    \centering
	\includegraphics[width=\linewidth]{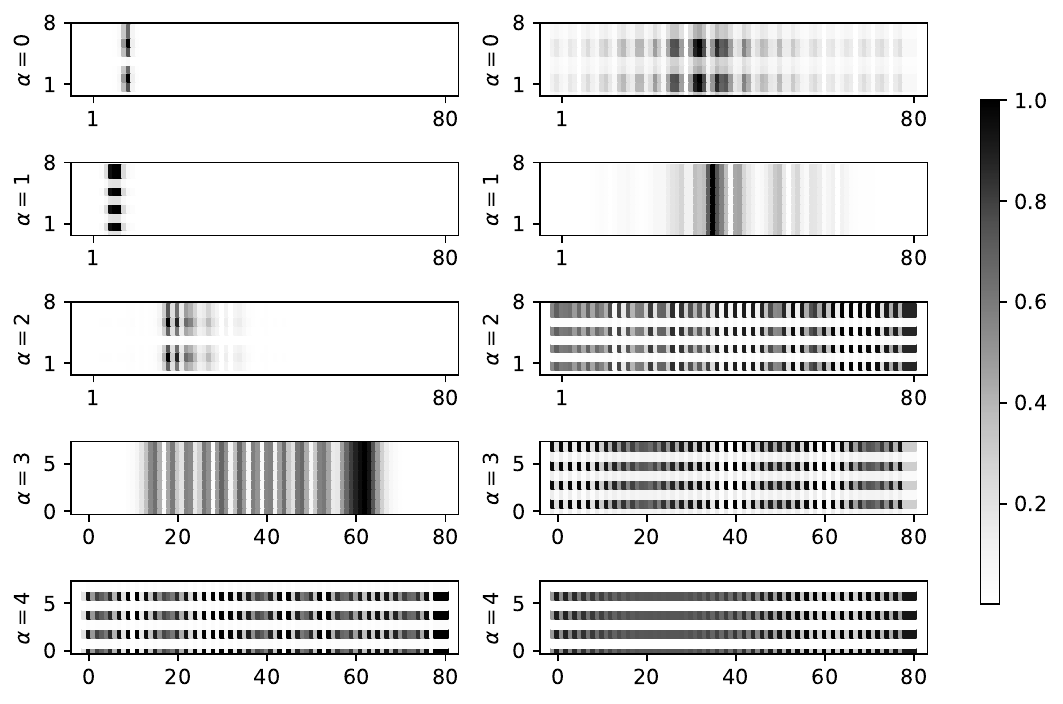}
\caption{Second example of absolute value of the wave-function configurations at the Fermi level ($E_F = 0$) for five different values of $\alpha$. Left column: row-wise randomness structure. Right column: fully correlated randomness structure. The wave-function values are normalized by the maximum value of the wave-function, scaled between 0 and 1.}
    \label{fig:s2}
\end{figure}

In Fig.~\ref{fig:s1},  and Fig.~\ref{fig:s2} we provide two examples of the wave-function configurations without any sample averaging. Each plot showcases the wave-function distribution for different values of $\alpha$. The left column shows the configurations for the row-wise randomness structure, and the right column shows the configurations for the fully correlated randomness structure.

These visual representations of the wave-function provide a qualitative understanding of how disorder and correlations affect the spatial distribution of the wave-function in a 2D tight-binding system. The examples highlight the differences in localization properties between row-wise and fully correlated randomness structures as the correlation parameter $\alpha$ varies.

We observe distinct changes in the distribution of wave-functions along rows and columns as $\alpha$ varies. This visual inspection underscores the need to quantify these behaviors rigorously. Single SPEE and SPRE provide robust measures to quantify the directional spread of the wave-function, emphasizing the importance of the cutting direction in these analyses.

\subsection{Numerical Results for SPEE and SPRE}
We analyze the SPEE and SPRE for various values of the correlation parameter $\alpha$, ranging from 0 to 4. Our observations are as follows.

\subsubsection{Row-Wise Disorder Results}
\begin{figure}
\centering
\begin{subfigure}{}%
\includegraphics[width=0.23\textwidth]{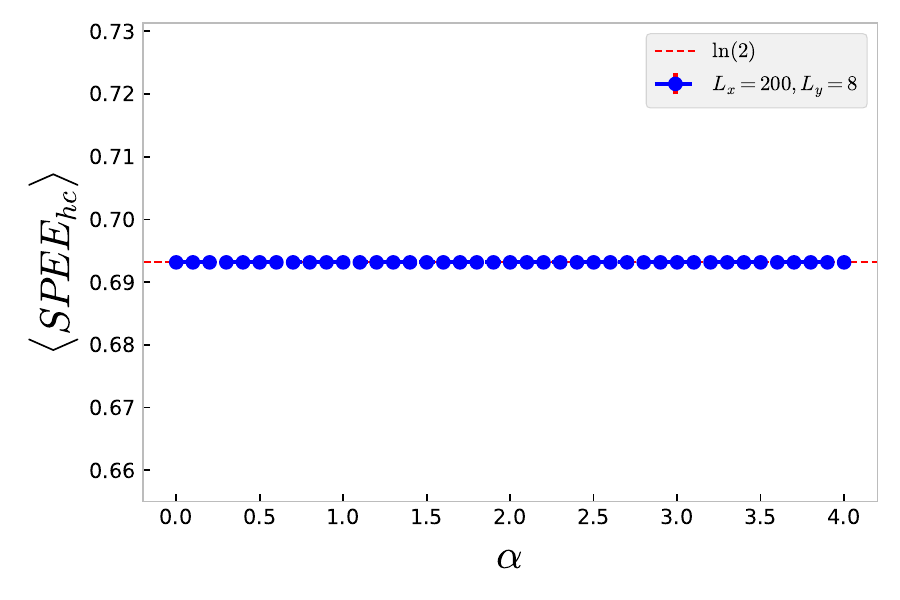}
\end{subfigure}
\begin{subfigure}{}%
\includegraphics[width=0.23\textwidth]{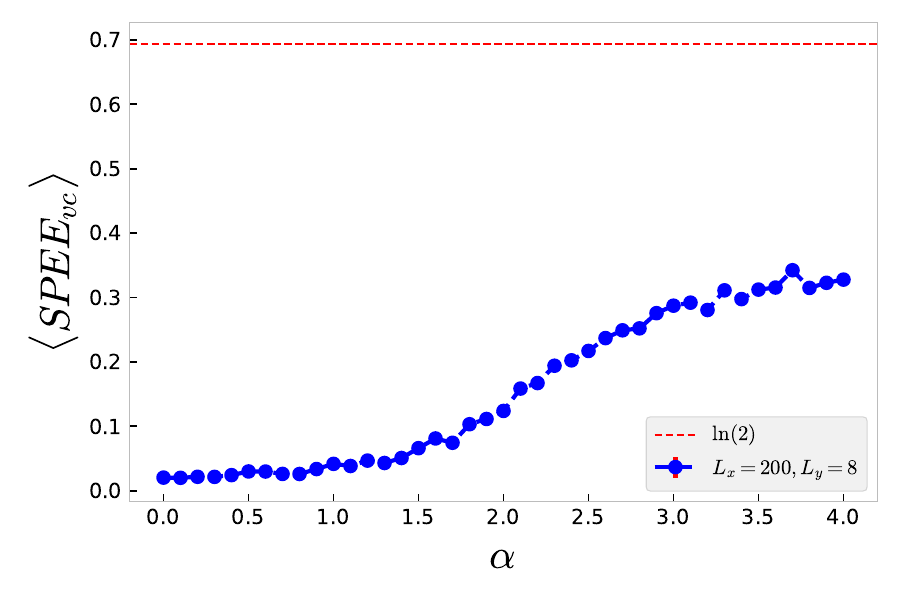}
\end{subfigure}%
\caption{Single Particle Entanglement Entropy (SPEE) for a 2D tight-binding system with row-wise disorder, averaged over $2000$ samples. Left: SPEE with horizontal cuts. Right: SPEE with vertical cuts. Each plot includes a horizontal line at $\ln(2)$ for reference. Error bars represent the standard error of the mean.\label{fig:speerow}}
\end{figure}

\begin{figure}
\centering
\begin{subfigure}{}%
\includegraphics[width=0.23\textwidth]{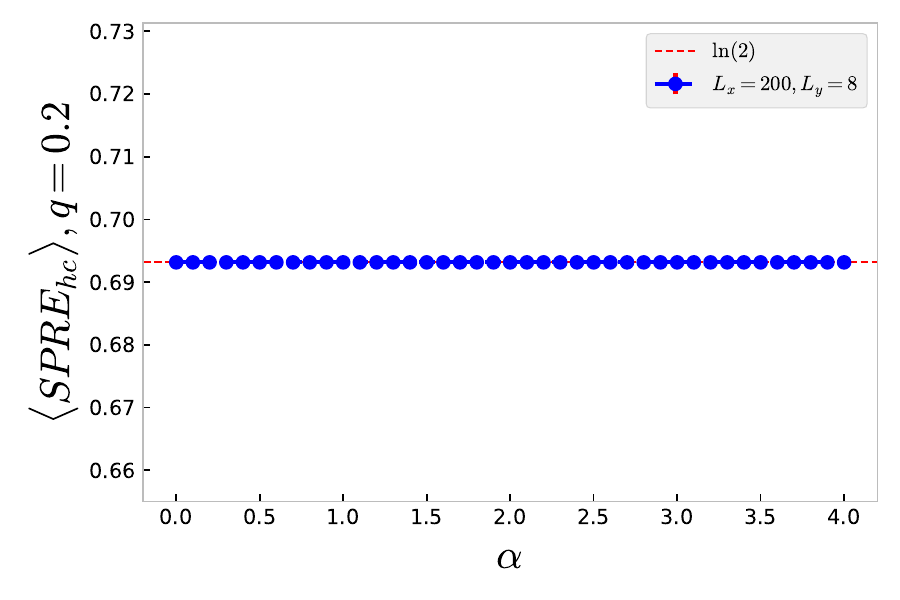}
\end{subfigure}
\begin{subfigure}{}%
\includegraphics[width=0.23\textwidth]{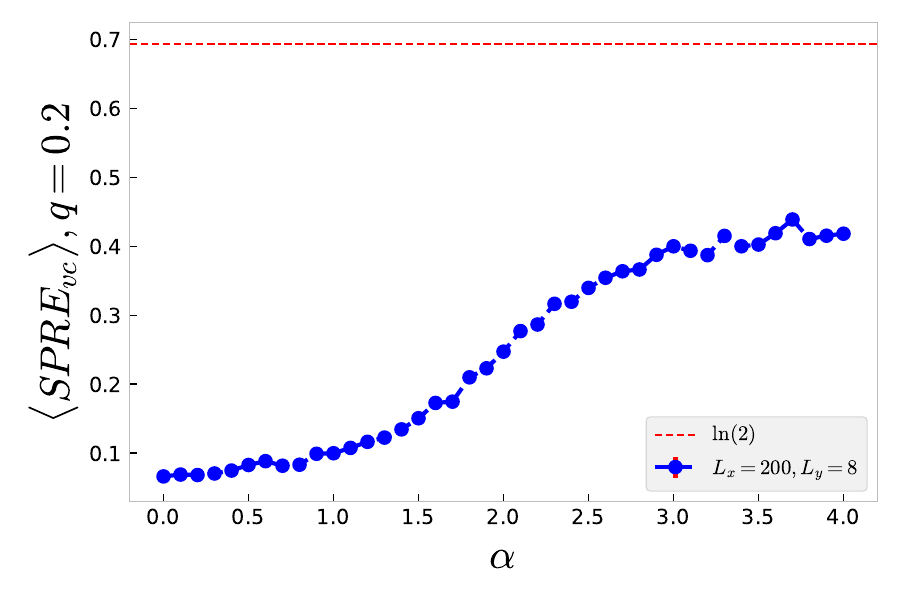}
\end{subfigure}%
\begin{subfigure}{}%
\includegraphics[width=0.23\textwidth]{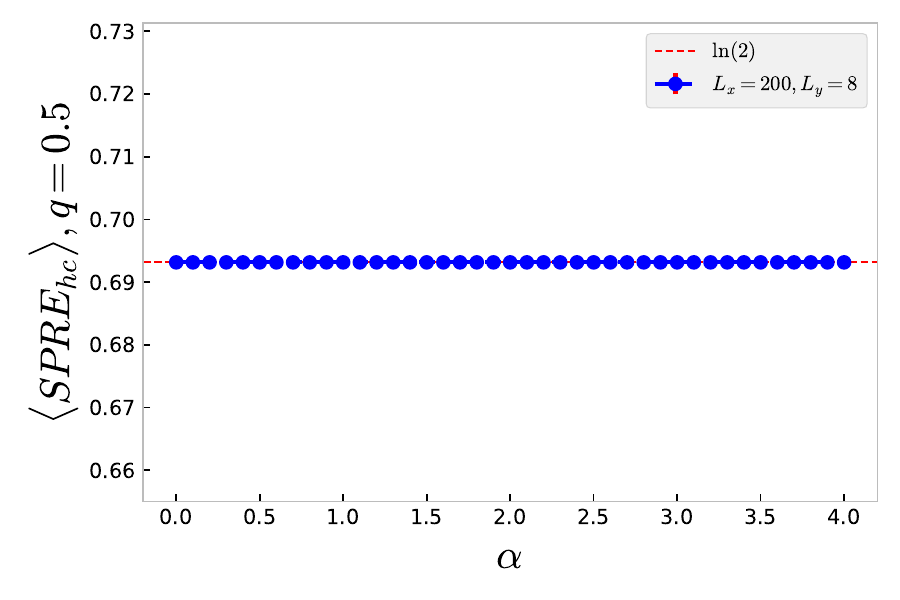}
\end{subfigure}
\begin{subfigure}{}%
\includegraphics[width=0.23\textwidth]{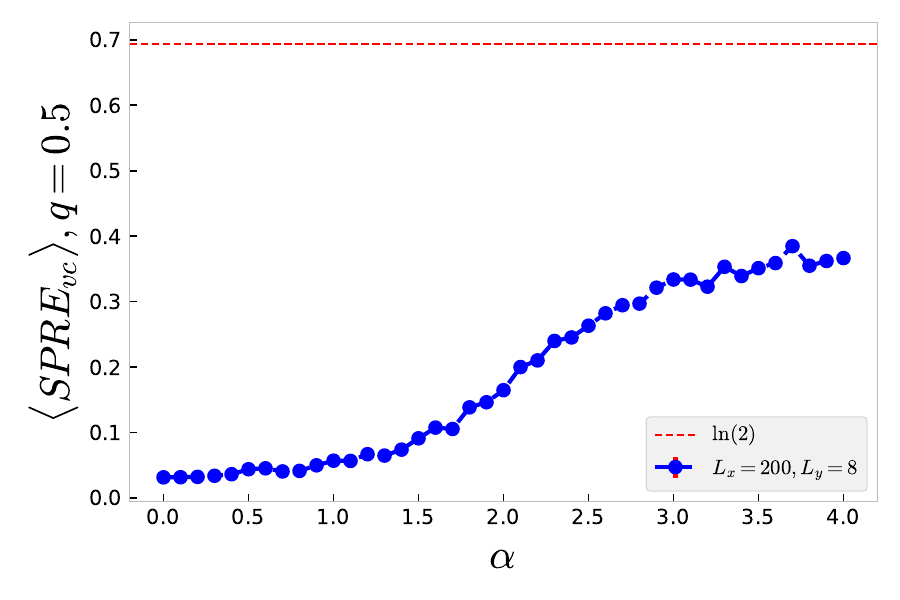}
\end{subfigure}%
\begin{subfigure}{}%
\includegraphics[width=0.23\textwidth]{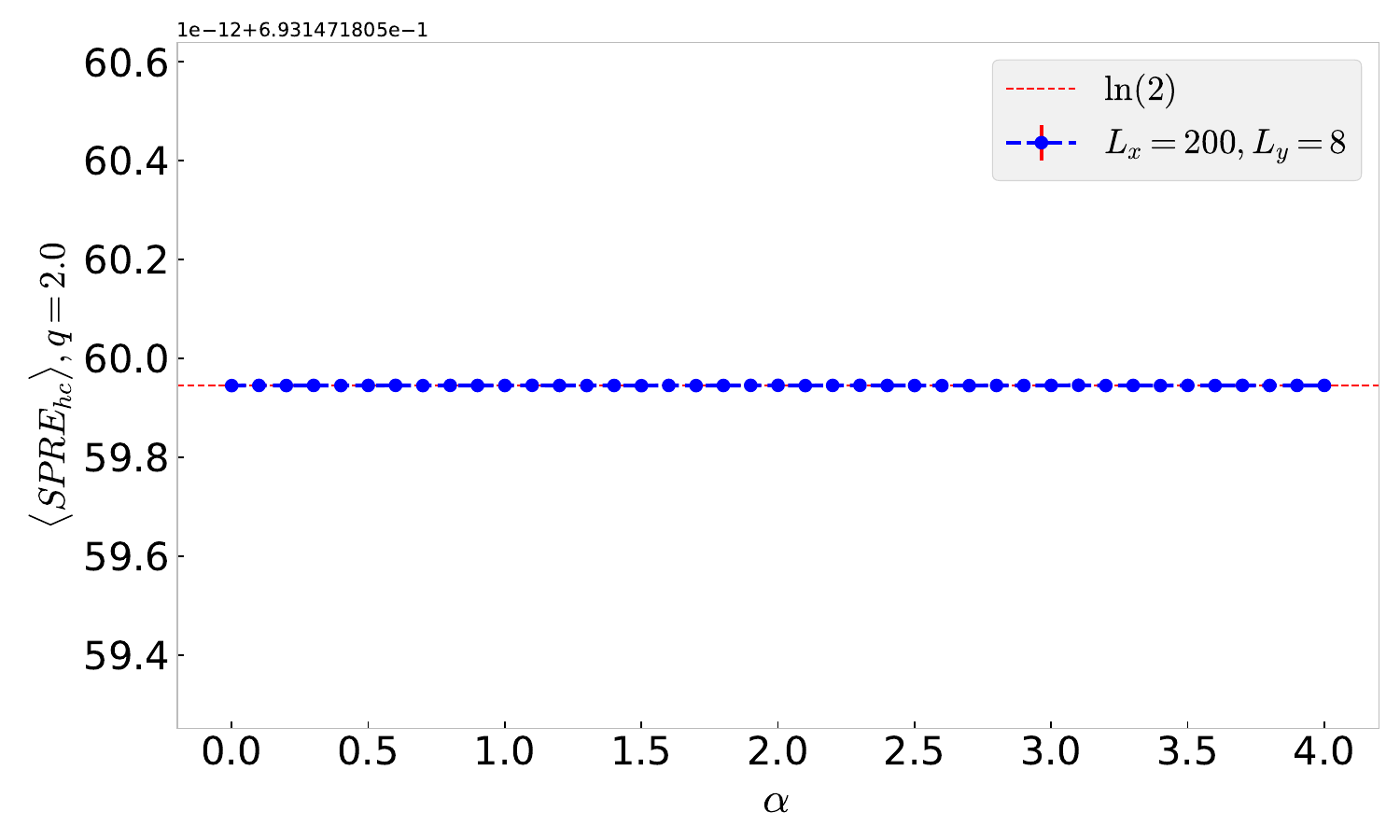}
\end{subfigure}
\begin{subfigure}{}%
\includegraphics[width=0.23\textwidth]{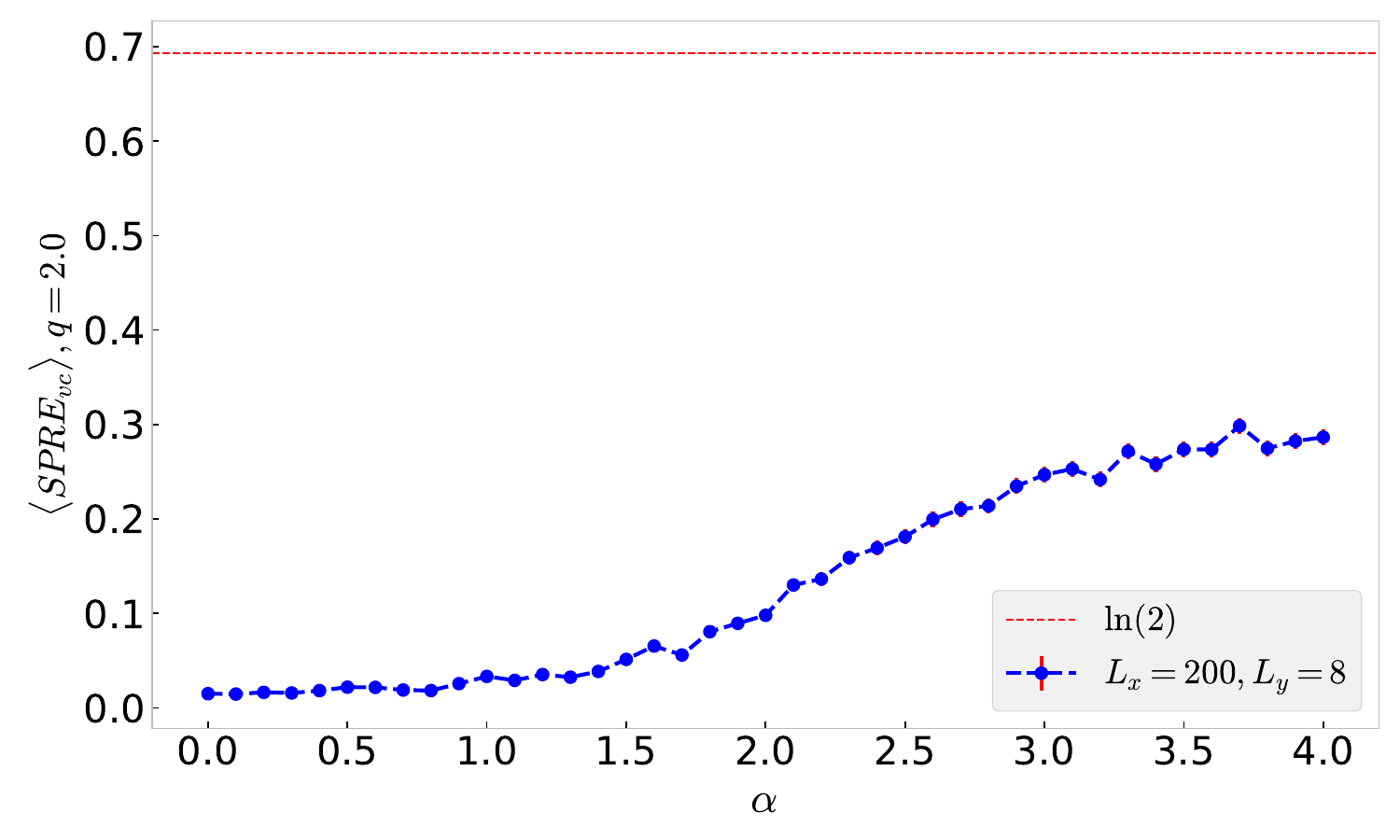}
\end{subfigure}%
\caption{Single Particle R\'enyi Entropy (SPRE) for a two-dimensional tight-binding system with row-wise disorder, averaged over $2000$ disorder realizations. The figure contains six subplots, organized in three rows and two columns. Each row corresponds to a different R\'enyi parameter: $q=0.2$ in the first row, $q=0.5$ in the second row, and $q=2$ in the third row. The left column shows the SPRE for horizontal cuts, while the right column displays the SPRE for vertical cuts. Each plot includes a reference horizontal line at $\ln(2)$ to guide the reader. Error bars represent the standard error of the mean.}
\label{fig:sprerow}
\end{figure}

In Figures \ref{fig:speerow} and \ref{fig:sprerow}, we analyze the SPEE and SPRE with row-wise disorder. Figure \ref{fig:speerow} shows SPEE with horizontal cuts (left) and vertical cuts (right), both including a reference horizontal line at $\ln(2)$. Figure \ref{fig:sprerow} presents the SPRE in a grid of six plots, arranged in three rows and two columns. Each row corresponds to a different R\'enyi parameter: $q=0.2$ (first row), $q=0.5$ (second row), and $q=2$ (third row). The left column depicts the SPRE for horizontal cuts, while the right column shows the SPRE for vertical cuts. A reference horizontal line at $\ln(2)$ is included in each plot.

For the row-wise randomness structure, we observe distinct behaviors for horizontal and vertical cuts. In the case of the horizontal cut, both the SPEE and SPRE remain constant at $\ln(2)$ across all values of $\alpha$, regardless of the R\'enyi parameter $q$. This consistency suggests that the wave-function's entanglement remains maximally distributed between the upper and lower halves of the lattice, showing no sensitivity to the correlation strength along this direction.

On the other hand, for the vertical cut, a different trend emerges. The SPEE and SPRE start from very small values (approaching zero) at low values of $\alpha$, and progressively increase as the correlation strength increases. Eventually, they saturate to a fixed value that depends on the system size but remains below $\ln(2)$. This behavior implies that, as the correlation strength increases, the wave-function becomes more extended along the horizontal direction, leading to greater entanglement between the left and right halves of the system.

Notably, the results for $ q = 2 $, linked to the inverse participation ratio (IPR) and purity, are shown in the third row of Figure \ref{fig:sprerow}. The SPRE at $ q = 2 $ provides insights into wave-function localization, where higher values indicate more delocalized states. The study of spatial-localization properties in disordered systems reveals significant connections between entropies and wave-function characteristics.

\subsubsection{Fully Correlated Randomness Results}

\begin{figure}
\centering
\begin{subfigure}{}%
\includegraphics[width=0.23\textwidth]{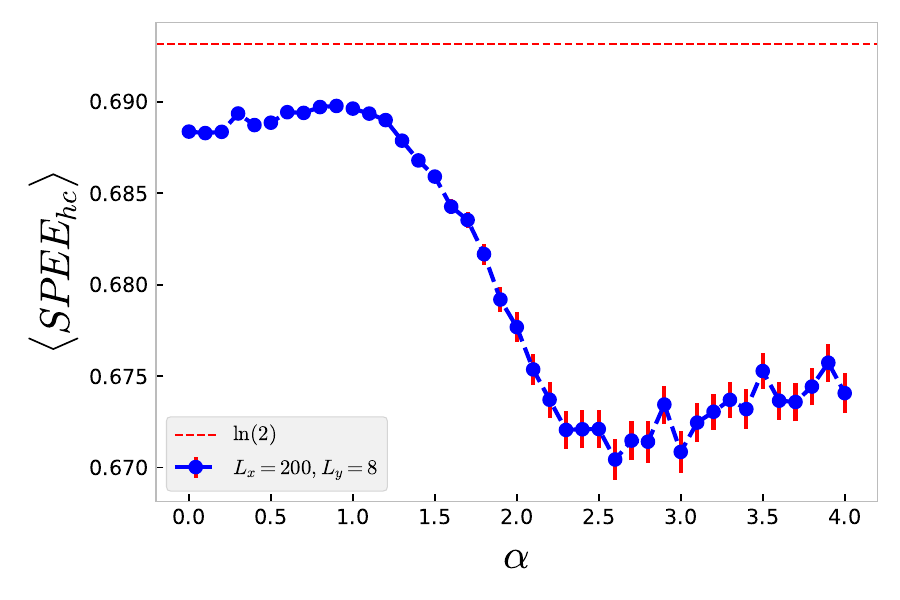}
\end{subfigure}
\begin{subfigure}{}%
\includegraphics[width=0.23\textwidth]{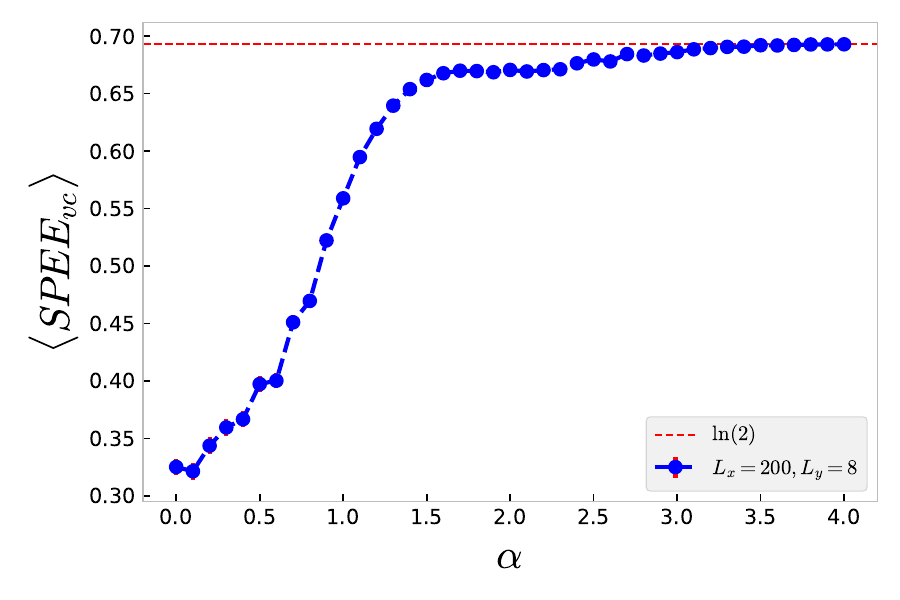}
\end{subfigure}%
\caption{Single Particle Entanglement Entropy (SPRE) for a 2D tight-binding system with fully correlated randomness, averaged over 2000 samples. Left: SPEE with horizontal cuts. Right: SPEE with vertical cuts. Each plot includes a horizontal line at $\ln(2)$ for reference. Error bars represent the standard error of the mean.\label{fig:speesame}}
\end{figure}

\begin{figure}
\centering
\begin{subfigure}{}%
\includegraphics[width=0.23\textwidth]{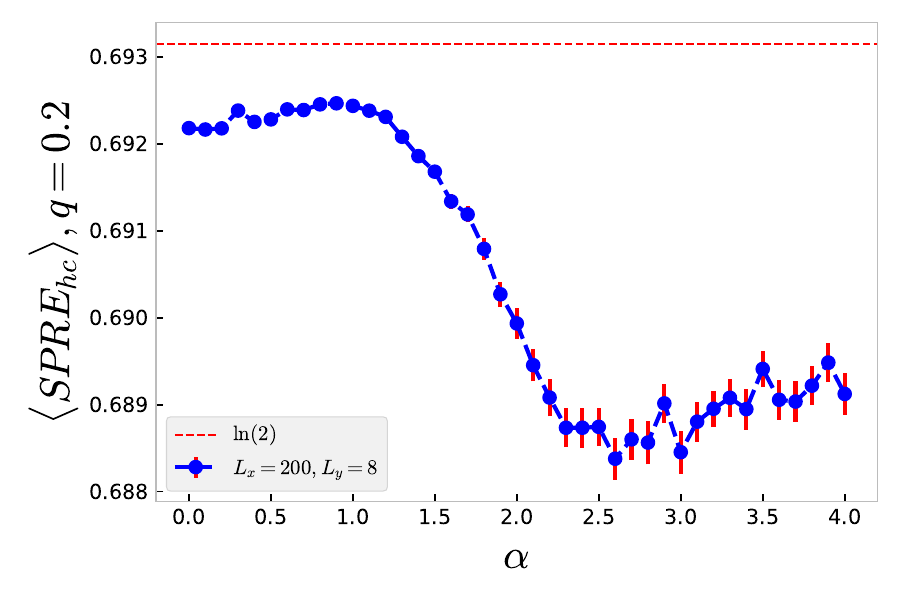}
\end{subfigure}
\begin{subfigure}{}%
\includegraphics[width=0.23\textwidth]{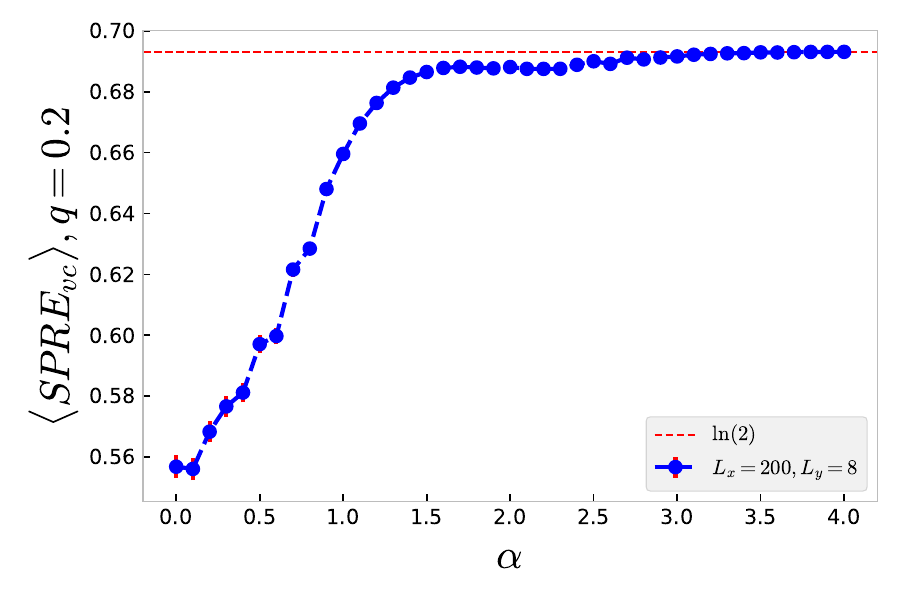}
\end{subfigure}%
\begin{subfigure}{}%
\includegraphics[width=0.23\textwidth]{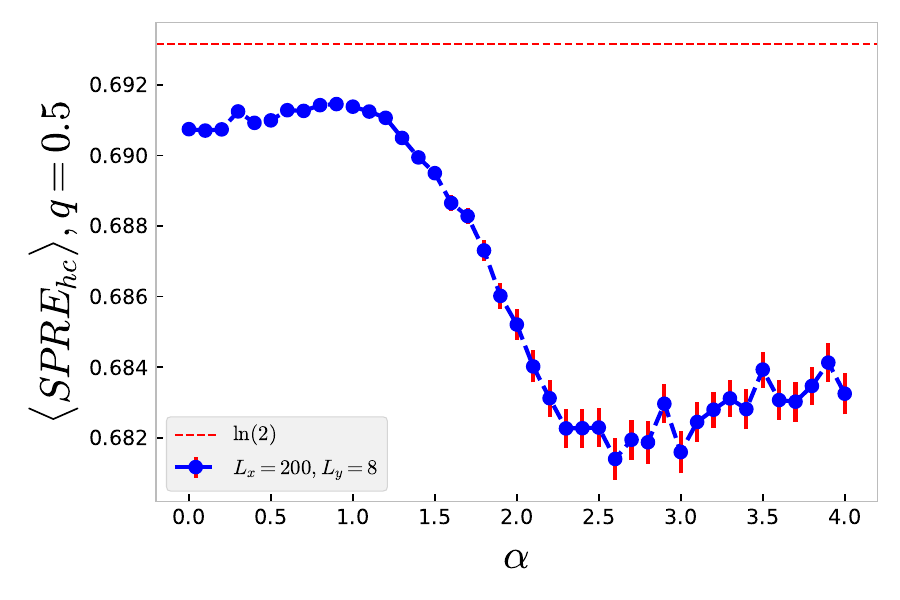}
\end{subfigure}
\begin{subfigure}{}%
\includegraphics[width=0.23\textwidth]{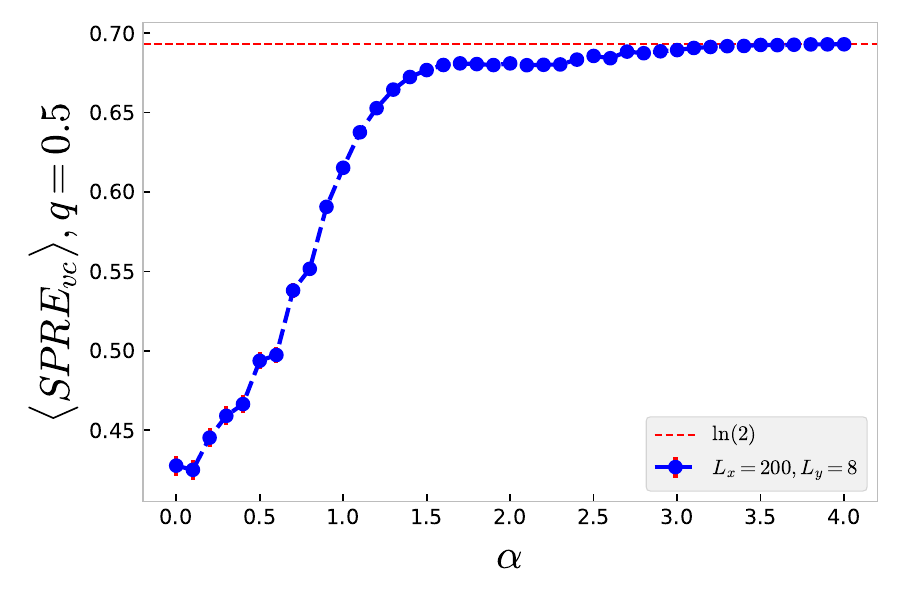}
\end{subfigure}%
\begin{subfigure}{}%
\includegraphics[width=0.23\textwidth]{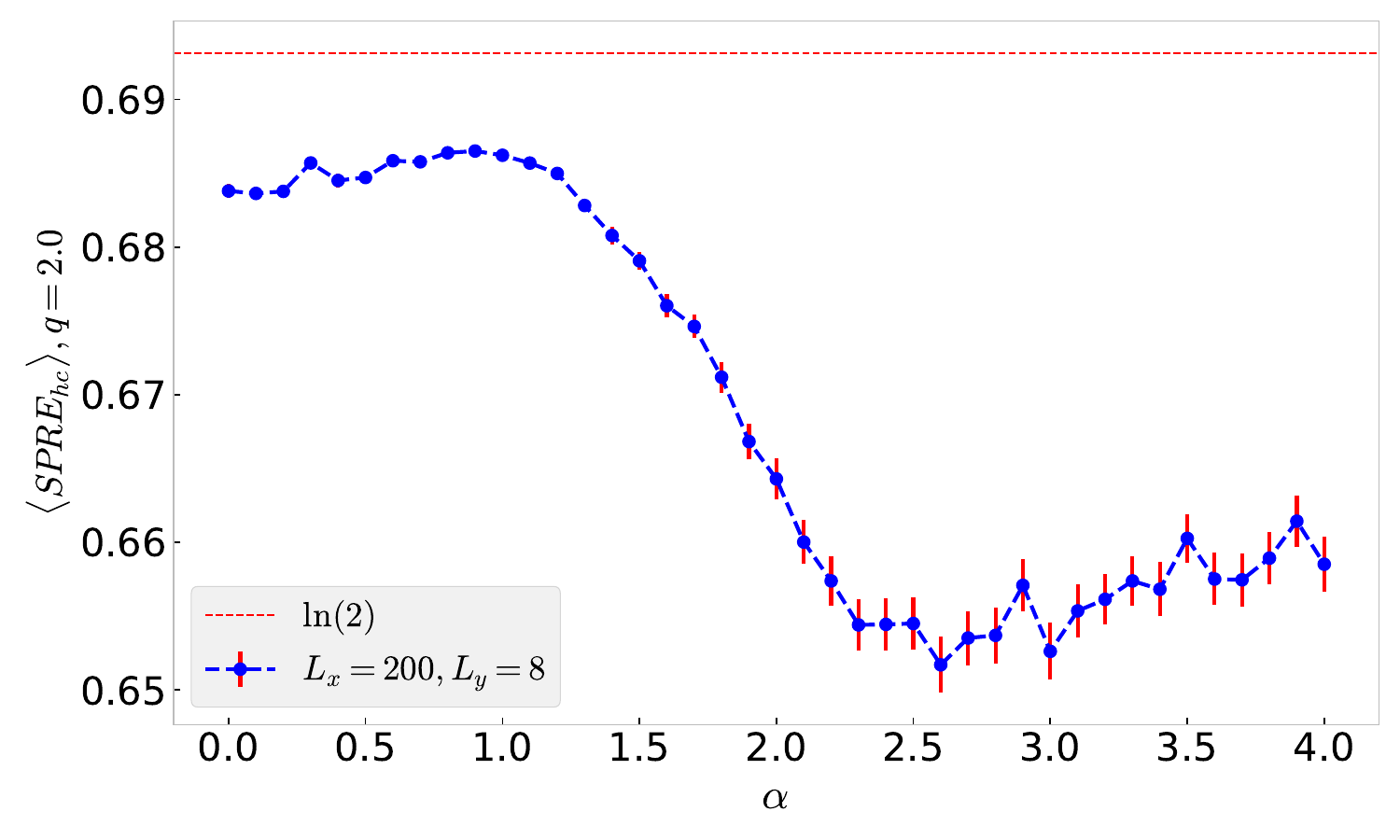}
\end{subfigure}
\begin{subfigure}{}%
\includegraphics[width=0.23\textwidth]{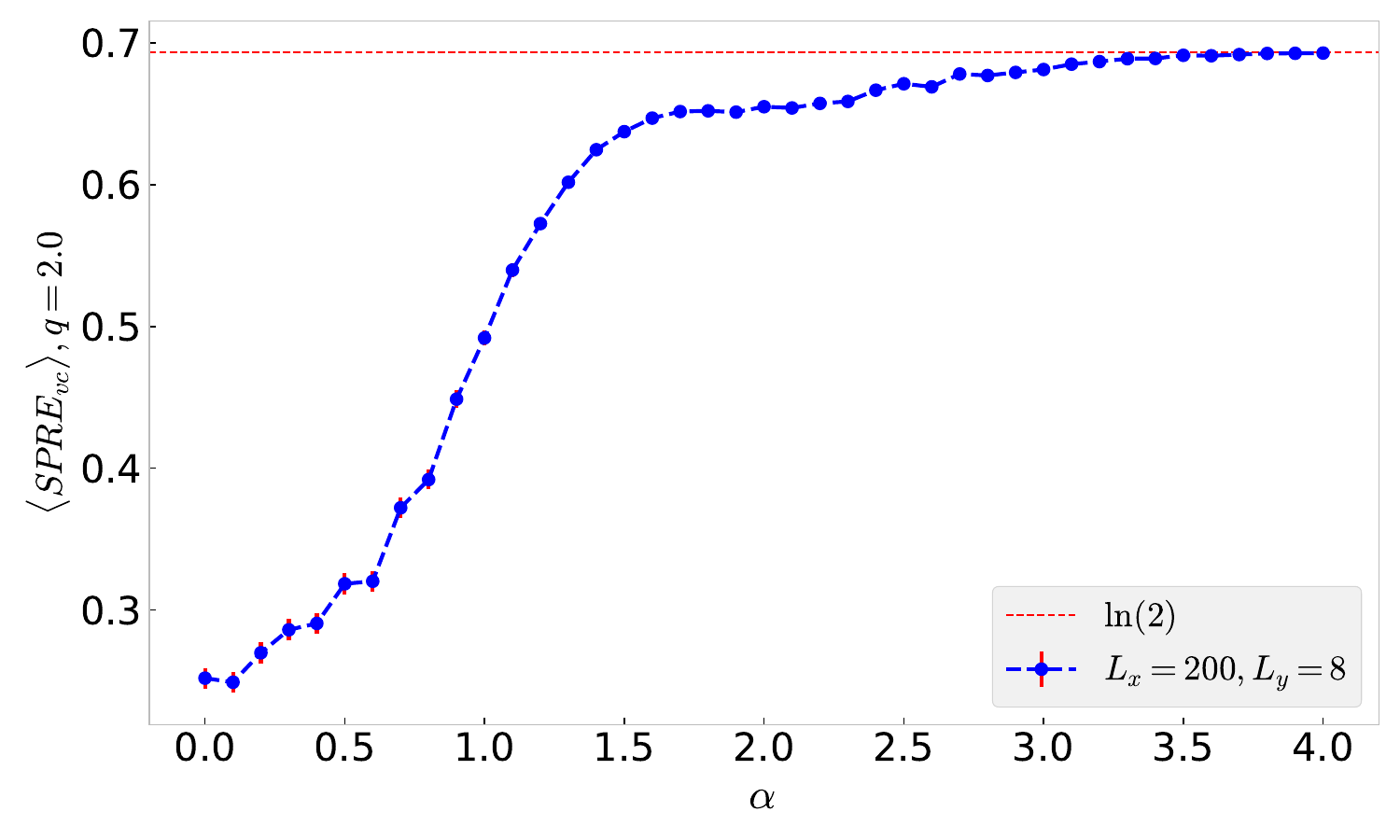}
\end{subfigure}%
\caption{Single Particle R\'enyi Entropy (SPRE) for a two-dimensional tight-binding system with fully correlated randomness, averaged over $2000$ disorder realizations. The figure consists of six subplots arranged in three rows and two columns. Each row corresponds to a different R\'enyi parameter: $q=0.2$ in the first row, $q=0.5$ in the second row, and $q=2$ in the third row. The left column presents the SPRE for horizontal cuts, while the right column displays the SPRE for vertical cuts. A reference horizontal line at $\ln(2)$ is included in each plot for guidance. Error bars represent the standard error of the mean.}
\label{fig:spresame}
\end{figure}

Figures \ref{fig:speesame} and \ref{fig:spresame} explore the SPEE and SPRE results with fully correlated randomness. In Figure \ref{fig:speesame}, SPEE is examined with horizontal cuts (left) and vertical cuts (right). Figure \ref{fig:spresame} presents SPRE with horizontal cuts (top row) and vertical cuts (bottom row) for $q=0.2$ (left) and $q=0.5$ (right).

For the fully correlated randomness structure, we observe the following. For the horizontal cut, the SPEE and SPRE begin slightly below $\ln(2)$ and decrease as $\alpha$ increases, ultimately saturating to a value dependent on the system size. This saturated value does not differ significantly from $\ln(2)$; for instance, it decreases from $0.69$ to $0.63$. This suggests that the wave-function's entanglement in the vertical direction diminishes slightly with increasing correlation. For $ q = 2 $, which follows a similar trend, the values also start from slightly below $\ln(2)$ and exhibit the same decreasing behavior.

Conversely, for the vertical cut, the SPEE and SPRE start from a fraction of $\ln(2)$ and increase with $\alpha$, eventually saturating at $\ln(2)$. This indicates that the wave-function becomes increasingly entangled along the vertical direction as the correlation strength rises, suggesting a delocalization trend in this direction. The behavior for $ q = 2 $ aligns with this observation, as it also reaches saturation at $\ln(2)$, highlighting the consistency across different R\'enyi parameters.

These results provide a detailed understanding of how the wave-function's entanglement properties vary with the correlation parameter $\alpha$ and how the direction of the cut influences the interpretation of SPEE and SPRE. The contrasting behaviors observed in horizontal and vertical cuts underscore the importance of considering directional properties in the study of localization and entanglement in disordered systems.

\subsubsection{Variations in R\'enyi Entropy with Changing $q$}
We investigate the relationships between the R\'enyi entropies as the parameter $ q $ varies, specifically examining the differences between entropies calculated at $ q = 0.2 $, $ q = 2 $, and the single-particle entanglement entropy (SPEE). We compute $ \text{SPRE}(q=0.2) - \text{SPRE}(q=2) $ and $ \text{SPEE} - \text{SPRE}(q=2) $. Our results indicate that both differences are consistently positive, aligning with the known inequality $ \text{SPRE}(0.2) > \text{SPRE}(0.5) > \text{SPRE}(q=2) = \text{SPEE} $ which can be attributed to the behavior of the probabilities $ p_A $ and $ p_B $ (associated with the subsystems), which are bounded between zero and one \cite{pipek1992universal, https://doi.org/10.1002/qua.560510619, PhysRevE.68.026202}. Furthermore, these differences clearly indicate a phase transition point at $ \alpha = 2 $, consistent with findings from previous studies. This analysis highlights the significance of examining R\'enyi entropy variations to gain deeper insights into the wave-function localization properties in the studied system.

\begin{figure}
\centering
\begin{subfigure}{}%
\includegraphics[width=0.23\textwidth]{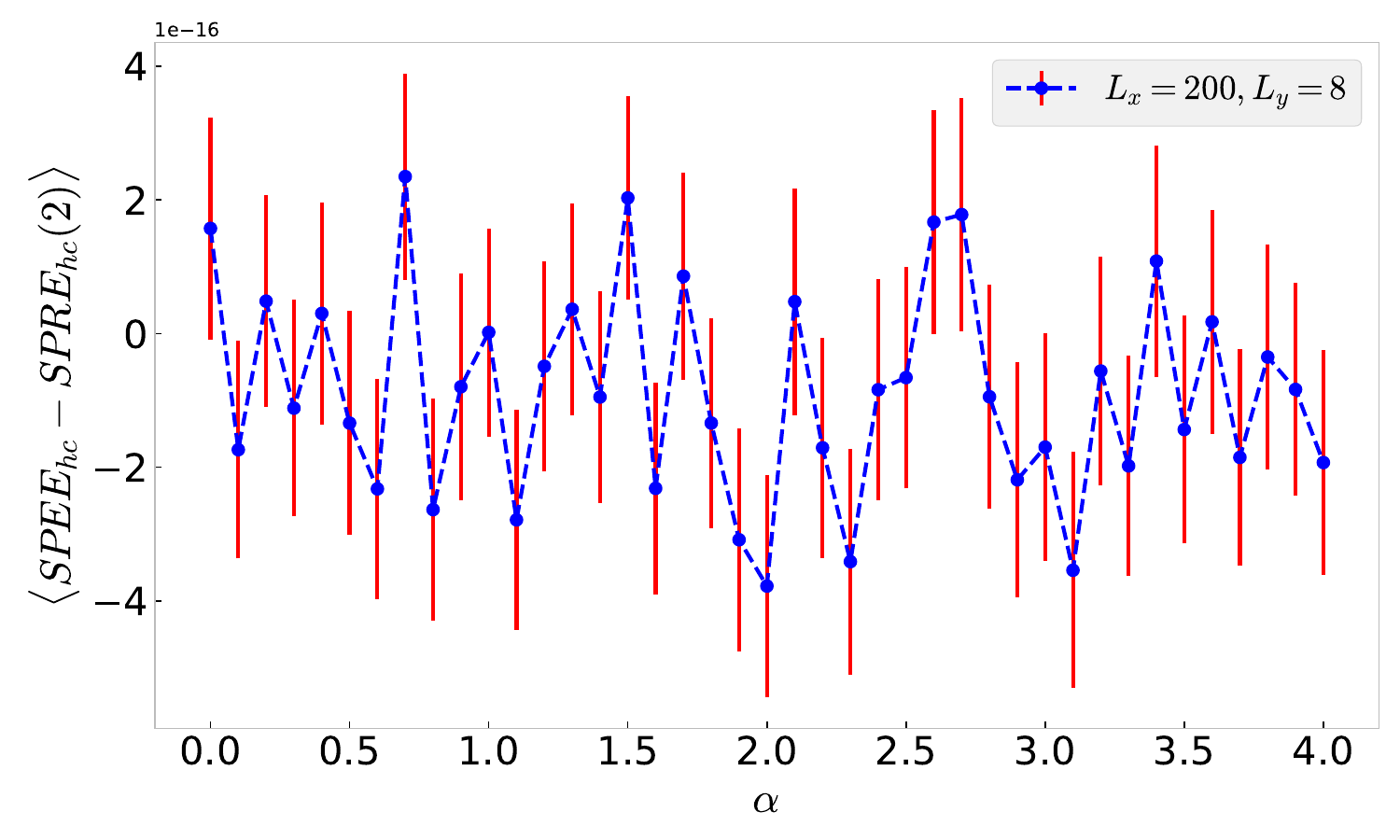}
\end{subfigure}
\begin{subfigure}{}%
\includegraphics[width=0.23\textwidth]{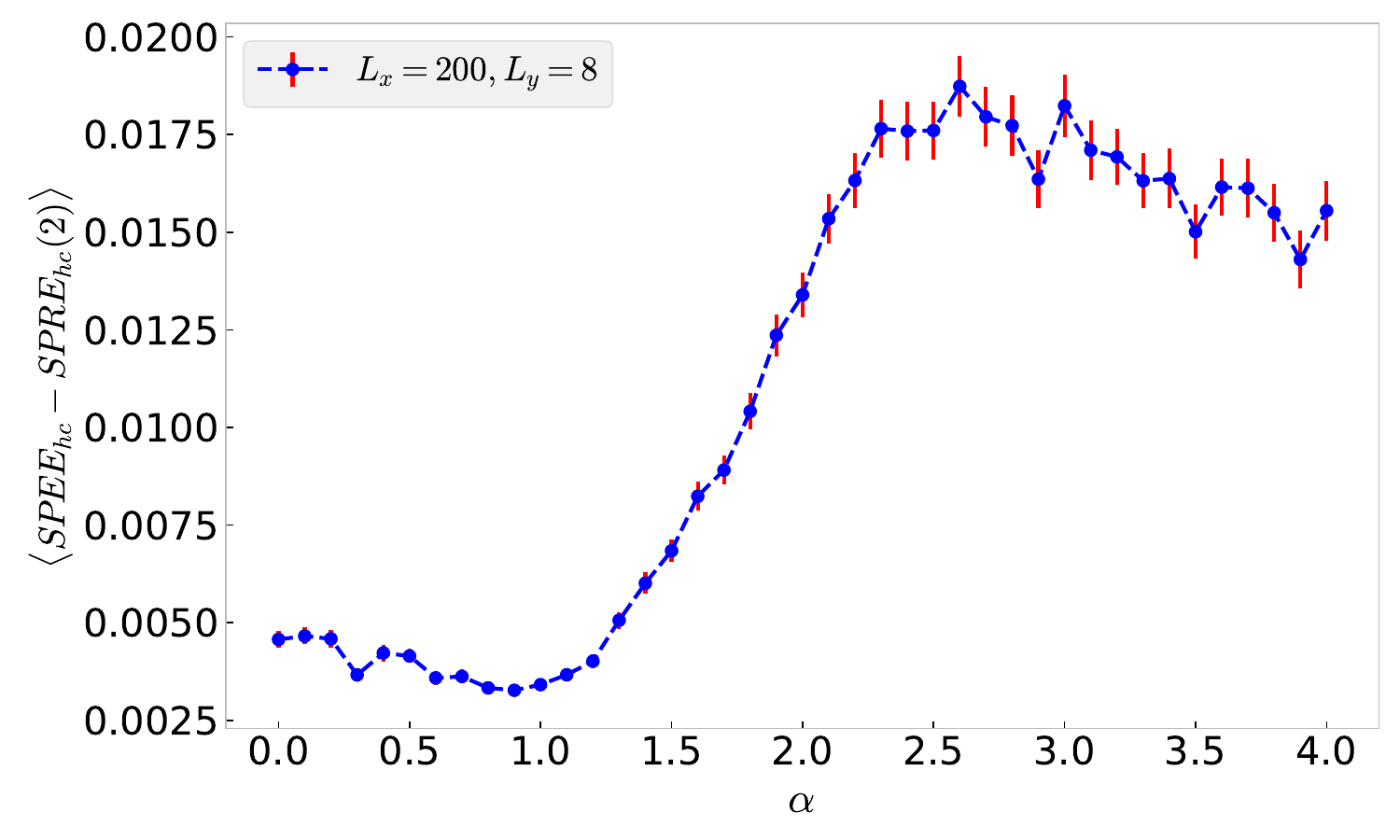}
\end{subfigure}%

\begin{subfigure}{}%
\includegraphics[width=0.23\textwidth]{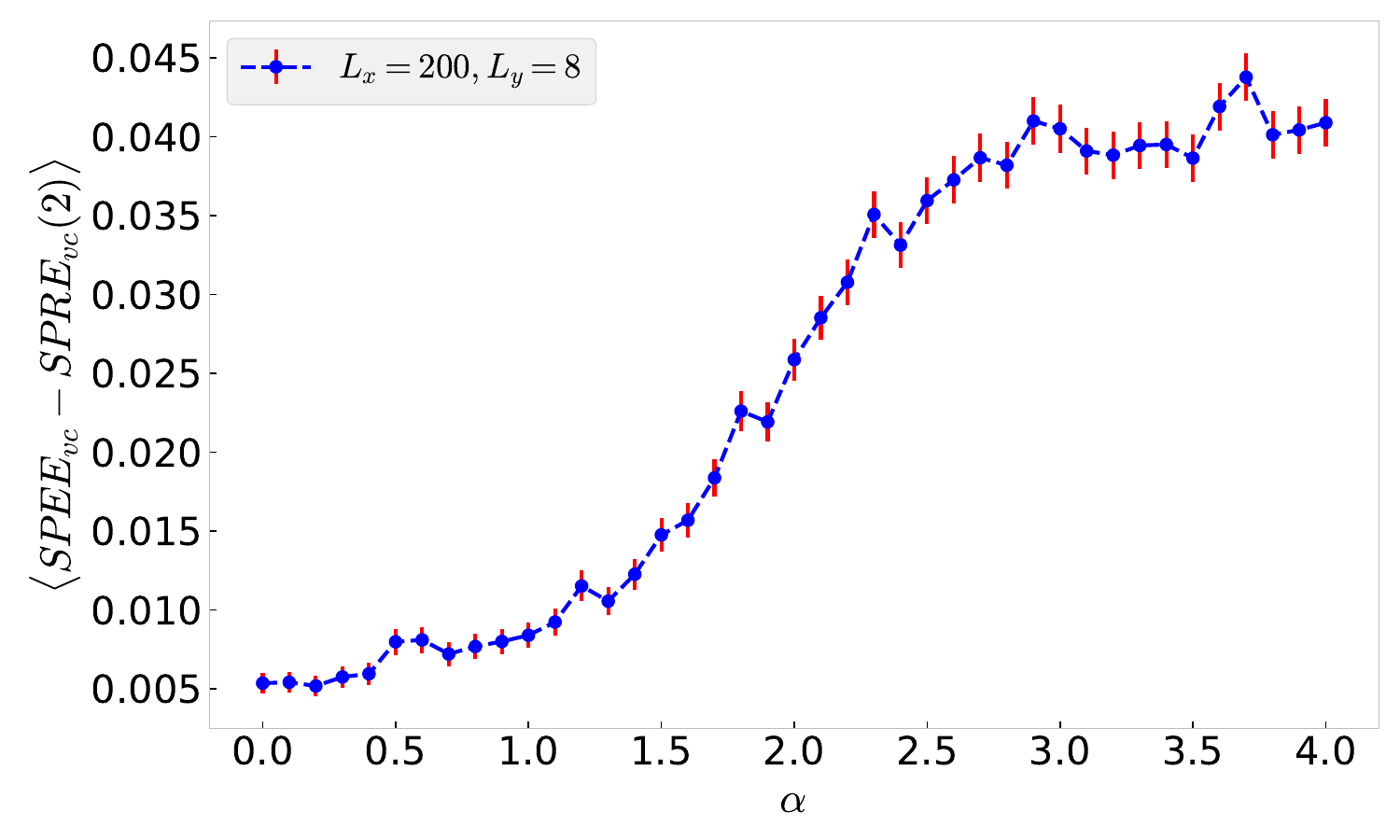}
\end{subfigure}%
\begin{subfigure}{}%
\includegraphics[width=0.23\textwidth]{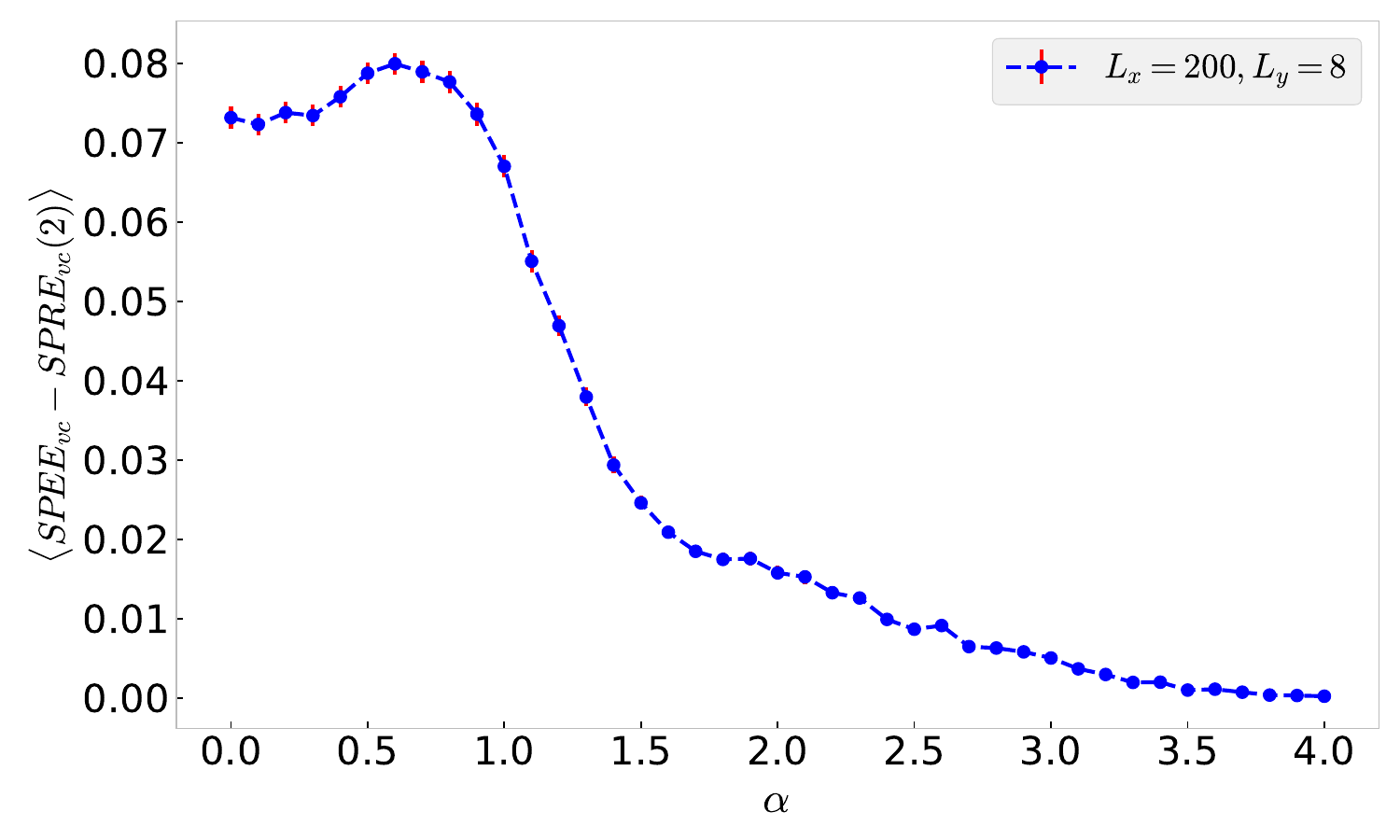}
\end{subfigure}%

\begin{subfigure}{}%
\includegraphics[width=0.23\textwidth]{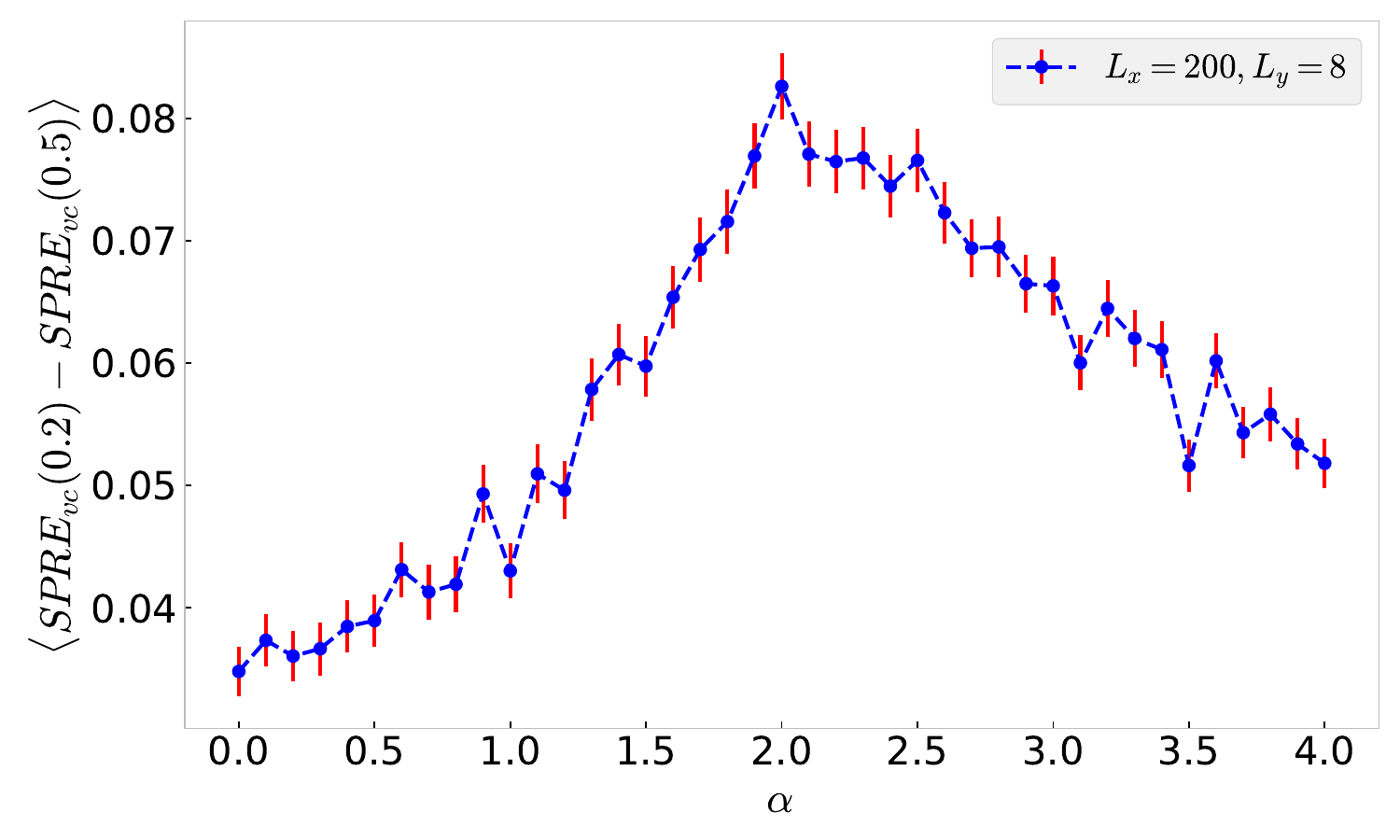}
\end{subfigure}
\begin{subfigure}{}%
\includegraphics[width=0.23\textwidth]{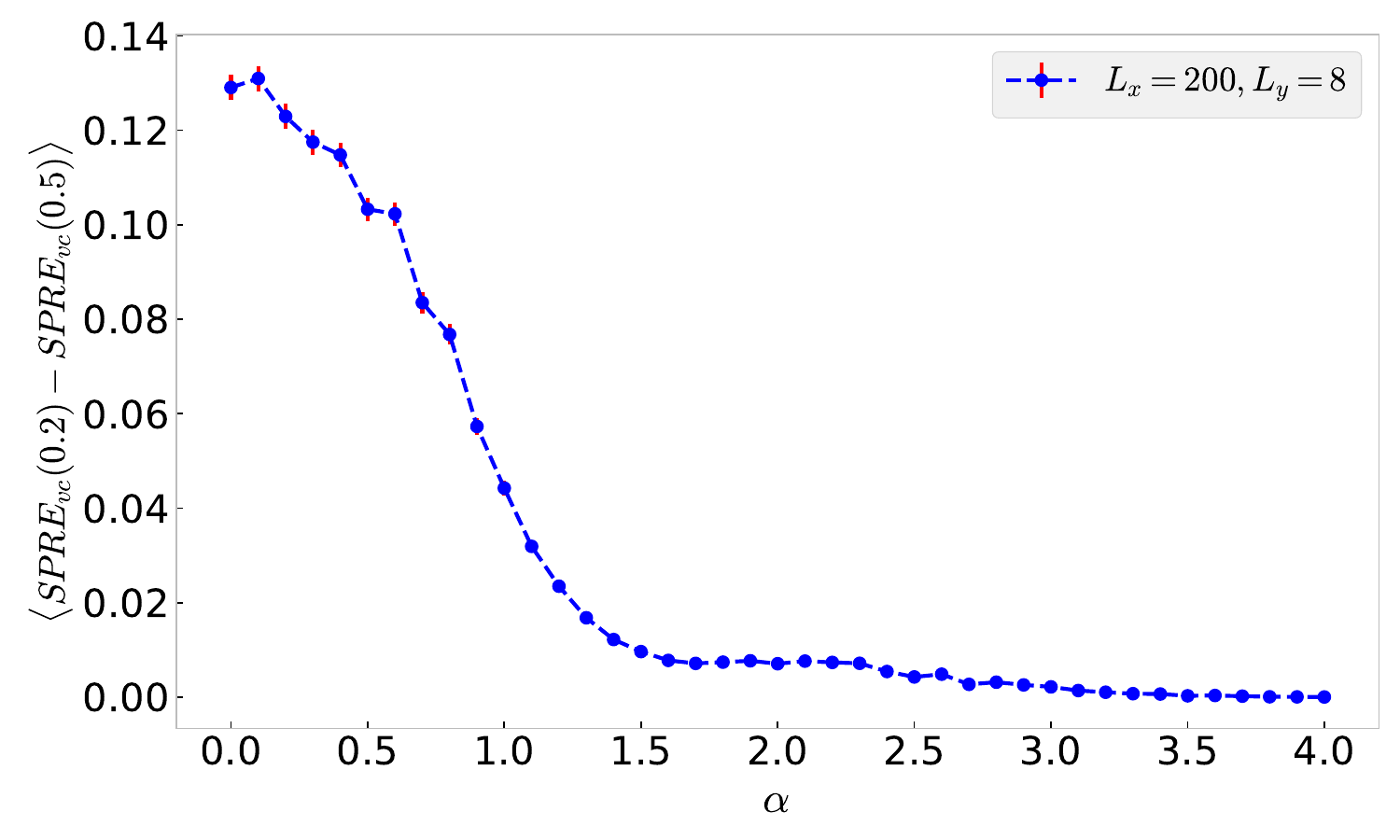}
\end{subfigure}%
\caption{Differences between R\'enyi entropy and the single-particle entanglement entropy (SPEE) for each plot. The left column corresponds to the row-wise correlated case, while the right column represents the fully correlated case. The results are averaged over $2000$ disorder realizations. The plots clearly indicate the phase transition point at $ \alpha = 2 $.}
\label{fig:changeq}
\end{figure}

\section{Conclusion}
\label{sec:conclusion}

In this study, we investigated the effects of correlated disorder on wave-function localization and anisotropy in a 2D tight-binding model with uniform hopping and random on-site energies. By employing Single Particle Entanglement Entropy (SPEE) and Single Particle R\'enyi Entropy (SPRE) for different values of $q$, we demonstrated that the disorder correlation parameter $\alpha$ plays a crucial role in shaping both the localization properties and directional spread of the wave-function. Our results revealed distinct behaviors between row-wise and fully correlated disorder, highlighting the directional dependence of entanglement and localization transitions. Additionally, the comparison between SPEE and SPRE provided further insights into the nature of wave-function localization.

For future work, we propose investigating the impact of disorder correlations on other physical properties, such as transport characteristics and spectral statistics. Furthermore, exploring the dynamical evolution of wave-functions in the presence of correlated disorder may uncover new phenomena with implications for condensed matter physics and material science.

\section*{Data Availability Statement}
The data that support the findings of this study are available from the corresponding author, upon reasonable request.

\acknowledgments

\bibliographystyle{apsrev4-2.bst}
\bibliography{/home/cms/Dropbox/physics/Bib/reference.bib}
\end{document}